\documentclass[revsymb,aps,prb,12pt]{revtex4}

\begin{document}
\title{A long-wave action of spin Hamiltonians and \\ the inverse problem
of the calculus of variations. }
\author{I.G. Bostrem}
\author{A.S. Ovchinnikov}
\author{R.F. Egorov}
\address{Department of Theoretical Physics, Ural State University, \\
620083, Lenin Ave. 51, Ekaterinburg, Russia}

\date{\today}
\begin{abstract}
 We suggest a method of derivation of the long-wave action of the model
spin Hamiltonians using the non-linear partial differential
equations of motions of the individual spins. According to the
Vainberg's theorem the set of these equations are (formal)
potential if the symmetry analysis for the Frechet derivatives of
the system is true. The case of Heisenberg (anti)ferromagnets is
considered. It is shown the functional whose stationary points are
described by the equations coincides with the long-wave action and
includes the non-trivial topological term (Berry phase).
\end{abstract}


\maketitle

\newpage {}

The most of rigorous results in the dynamics of a one- or
two-dimensional isotropic Heisenberg antiferromagnet due to a presence of
non-trivial topological term in the action (Berry phase) has been obtained in a coherent state path integral formulation \cite{haldan}. This approach starts with the derivation a path integral representation for the partition function of an individual spin $Z= {Tr} \exp(-\beta H(\vec S))$ where the Hamiltonian $H(\vec S)$ is a linear function of the spin operator $\vec S$. Further, this result is generalized for a lattice problem using the fact the Hilbert space of the many spin system is the direct product of the one-spin Hilbert  spaces at each site.
In path-integral formulation is convenient to use an overcomplete basis of spin coherent states $|{\vec N} \rangle$  labeled by the points $\vec N$ on the surface of the unit sphere with the expectation value of the spin $\langle {\vec N}| \vec S |{\vec N} \rangle =S \vec N$.
The exponential in the expression for $Z$ is written as the Trotter product of a large number of exponentials each evolving the system over an infinitisimal imaginary time interval $\Delta \tau$ and using the resolution of unity $1=\int \frac{d {\vec N}}{2 \pi} |{\vec N} \rangle \langle {\vec N}|$ between the intervals (the integral is over the unit sphere). This procedure leads to the following representation for the partition function
\begin{equation}
Z=\int {D} \vec{N}(\tau )\exp \left[ \int\limits_{0}^{\beta }d\tau \left[
\left\langle \vec{N}(\tau )\left| \frac{d}{d\tau }\right| \vec{N}(\tau
)\right\rangle -H\left( S\vec{N}(\tau )\right) \right] \right], \label{eq:STAT}
\end{equation}
where $H ( S\vec{N}(\tau ) )$ is the Hamiltonian in spin coherent representation. The first term in the action is the Berry phase term $S_{{B}}$ and represents the overlap between the coherent states at two infinitesimally separated times. The manipulation $S_{{B}}$ into a physically  more transparent form  may be found, for example, in Ref. \cite{Sachdev}. The result is
\begin{equation}
S_{{B}}= i S \int\limits_{0}^{\beta } d {\tau} \int\limits_{0}^{1}
d u \left( \vec{N}\left[ \frac{\partial \vec{N}}{\partial u}\times
\frac{\partial \vec{N}}{\partial \tau }\right] \right),
\label{eq:BERRY}
\end{equation}
where $u$ is a dummy integration variable, $\left| \vec{N}(\tau ,u)\right\rangle$ is defined by
\begin{equation}
\left| \vec{N}(\tau ,u)\right\rangle =\exp \left( u\left( z(\tau )\hat{S}%
^{+}-z^{*}(\tau )\hat{S}^{-}\right) \right) \left| \vec{N}%
=(0,0,1)\right\rangle,
\end{equation}
with the properties $\left| \vec{N}(\tau ,u=1)\right\rangle =\left| \vec{N}(\tau )\right\rangle $ , $\left| \vec{N}(\tau ,u=0)\right\rangle =\left| \vec{N} = (0,0,1) \right\rangle $ and the complex number $z$ is given in spherical coordinates $z=-\frac{\Theta}{2}\exp(-{i} \phi)$.
The  $S_{{B}}$ presents the oriented area contained within the closed loop defined by $\vec{N}(\tau)$, $0 \leq \tau \leq \beta$, $\vec{N}(0)=\vec{N}(\beta)$  on the spin sphere.

In this Letter, it will be demonstrated how the long-wave action for spin Hamiltonians can be derived by using the non-linear differential equations of motion of spins via the solving of the inverse problem of the calculus of variations: that is, finding the functional whose stationary points are described by the descriptive equations.

As is well known the most fundamental question is whether the functional
exists for a given operator. In the Ref.\cite{Atherton} Atherton and Homsy
have suggested a stringent and elegant formalism based on Vainberg's theorem
\cite{Vainberg} to derive the operational formulas for the case of an
arbitrary number of nonlinear differential equations in an arbitrary number
of independent variables and of arbitrary order to determine whether given
differential operators are potential.

As the simplest illustration of the method we consider the case of a
Heisenberg ferromagnet with the spin Hamiltonian $H=-\frac{1}{2}\sum\limits_{\left\langle im\right\rangle}J_{im}\vec{S}_{i}\vec{S}_{m}$
where the $\vec{S}_{i}$ are spin $S$ quantum spin operators on the $i$th
sites of a $n$-dimensional hypercubic lattice. The exchange constant $J_{im}>0$ is non-zero just for the nearest
neighbors. The non-linear differential equations describing the dynamics of
Heisenberg ferromagnet can be obtained by taking the diagonal matrix
elements of the equation of motion for the raising operator
${i} \hbar \frac{\partial S_{i}^{+}}{\partial t}=[S_{i}^{+},H]$ of the $i$th spin in
spin-coherent representation $\left| \Omega \right\rangle
=\prod\limits_{i=1}^{N}\left| \theta _{i},\phi _{i}\right\rangle ,$ where
$0\leq \theta _{i}\leq \pi $ and $0\leq \phi _{i}<2\pi $ parametrize the spin
states on the unit sphere \cite{Bala}. In continuum approximation the
coupled c-number equations for the real variables $\theta $ and $\phi $
depending on the time $t\in [t_{1},t_{2}]$ and space coordinates $\vec{r}$ $\in
{\Re}^{n}$ can be written as following (hereafter the lattice constant is unit)
\[
0=-S\sin \Theta \frac{\partial \phi }{\partial t}+\frac{JS^{2}}{\hbar }
\left( \Delta \Theta -\cos \Theta \sin \Theta \left( \vec{\nabla}\phi
\right) ^{2}\right) ,
\]
\begin{equation}
0=S\frac{\partial \Theta }{\partial t}+\frac{JS^{2}}{\hbar }\left( 2\cos
\Theta \left( \vec{\nabla}\phi \vec{\nabla}\Theta \right) +\sin \Theta
\Delta \phi \right) .  \label{eq:EM2}
\end{equation}
\smallskip These equations present a vector of differential operators
${R}^{j}(z_{k})=0$ $\;(j, k = 1,2$ and $z_{1}=\Theta ,z_{2}=\phi )$
\[
{R}^{1}=-S\sin \Theta \frac{\partial \phi }{\partial t}+\frac{JS^{2}}{\hbar }
\left( \Delta \Theta -\cos \Theta \sin \Theta \left( \vec{\nabla}\phi
\right) ^{2}\right) ,
\]
\begin{equation}
{R}^{2}=S\sin \Theta \frac{\partial \Theta }{\partial t}+\frac{JS^{2}}{\hbar }
\left( 2\sin \Theta \cos \Theta \left( \vec{\nabla}\phi \vec{\nabla}\Theta
\right) +\sin ^{2}\Theta \Delta \phi \right) .  \label{eq:Rvec}
\end{equation}
The second equation in (\ref{eq:EM2}) is multiplied by the factor $\sin \Theta $
that is necessary for the further symmetry analysis. The inverse problem is
solved if an operator can be proven to be a potential operator, i.e. an
analysis of the Frechet derivatives is symmetrical (Vainberg's theorem).
According to the procedure in \cite{Atherton} we evaluate the "two-tensor" of
Frechet derivatives in the directions $u_{k}$
\begin{equation}
{R}_{,k}^{j}=\lim\limits_{\epsilon \rightarrow 0}\frac{1}{\epsilon }
\left( {R}^{j}(z_{k}+\epsilon u_{k})-{R}^{j}(z_{k})\right) ,  \label{eq:FD}
\end{equation}
\begin{equation}
{R}_{,1}^{1}=-u_{1}S\cos \Theta \frac{\partial \phi }{\partial t}+\frac{JS^{2}
}{\hbar }\left( \left( \Delta u_{1}\right) -u_{1}\cos 2\Theta \left( \vec{
\nabla}\phi \right) ^{2}\right) ,  \label{eq:FD11}
\end{equation}
\begin{equation}
{R}_{,2}^{1}=-S\sin \Theta \frac{\partial u_{2}}{\partial t}-\frac{JS^{2}}{
\hbar }\sin 2\Theta \left( \vec{\nabla}\phi \vec{\nabla}u_{2}\right) ,
\label{eq:FD12}
\end{equation}
\begin{eqnarray}
{R}_{,1}^{2} &=&S\sin \Theta \frac{\partial u_{1}}{\partial t}+u_{1}S\cos
\Theta \frac{\partial \Theta }{\partial t}  \nonumber \\
&&+\frac{JS^{2}}{\hbar }\left( u_{1}\sin 2\Theta \Delta \phi +\sin 2\Theta
\left( \vec{\nabla}u_{1}\vec{\nabla}\phi \right) +2u_{1}\cos 2\Theta \left(
\vec{\nabla}\phi \vec{\nabla}\Theta \right) \right) ,  \label{eq:FD21}
\end{eqnarray}
\begin{equation}
{R}_{,2}^{2}=\frac{JS^{2}}{\hbar }\sin ^{2}\Theta \Delta u_{2}+\frac{JS^{2}}{
\hbar }\sin 2\Theta \left( \vec{\nabla}u_{2}\vec{\nabla}\Theta \right) .
\label{eq:FD22}
\end{equation}
The symmetry requirement is satisfied if $\int
{\psi}_{j}{R}_{,k}^{j}u_{k}{d}V=\int
u_{j}{R}_{,k}^{j}{\psi}_{k}{d}V$ \smallskip for every
${\psi}_{j}$, $u_{k}$ in the range of the ${R}^{j}$. The integral
is taken over all space coordinates and time $\int {d}V\equiv
\int\limits_{t_{1}}^{t_{2}}{d}t\int {d}\vec{r}$. The respective
ordering of the vectors ${R}^{j}$ and $z_{k}$ is very important.
It determines the form of the variational integral and required
symmetry conditions.  The symmetry test yields to the following
boundary conditions
\begin{equation}
\int {\psi}_{1}{R}_{,1}^{1}u_{1}{d}V=\int
u_{1}{R}_{,1}^{1}{\psi}_{1}{d}V\smallskip \Rightarrow
\oint\limits_{S}({\psi}_{1}\vec{\nabla} u_{1}-u_{1}\vec{\nabla}
{\psi}_{1}){d}\vec{S}=0,  \label{eq:SM11}
\end{equation}
\begin{equation}
\int {\psi}_{1}{R}_{,2}^{1}u_{2}{d}V=\int
u_{2}{R}_{,1}^{2}{\psi}_{1}{d}V\smallskip \Rightarrow
\oint\limits_{S}\sin 2\Theta {\psi}_{1}u_{2}\vec{\nabla}\phi {d}
\vec{S}=0,  \smallskip {\psi}_{1}u_{2}\sin \Theta
|_{t_{2}}^{t_{1}}=0, \label{eq:SM12}
\end{equation}
\begin{equation}
\int {\psi}_{2}{R}_{,2}^{2}u_{2}{d}V=\int
u_{2}{R}_{,2}^{2}{\psi}_{2}{d}V\smallskip \Rightarrow
\oint\limits_{S}\sin ^{2}\Theta ({\psi}_{2}\vec{\nabla}
u_{2}-u_{2}\vec{\nabla} {\psi}_{2}){d}\vec{S}=0.  \label{eq:SM22}
\end{equation}

The variational principle for the differential system can be
constructed via the formula $\int
{d}V\int\limits_{0}^{1}{d}{\lambda}
\;z_{i}{R}^{i}({\lambda}z_{i})$ where
$\int\limits_{0}^{1}{d}{\lambda} $ represents integration over the
scalar variable $\lambda$. It results in the following
\begin{eqnarray}
F_{1} &=&\int {d}V\int\limits_{0}^{1} {d}{\lambda}\enskip \Theta
{R}_{1}({\lambda}\Theta )=\int {d}V\int\limits_{0}^{1}{d}{\lambda}
(-S\sin ({\lambda} \Theta ))\frac{
\partial ({\lambda} \Theta )}{\partial {\lambda}}\frac{\partial \phi }{\partial
t}  \nonumber \\
&&-\frac{JS^{2}}{2\hbar }\int {d}V\left( \sin ^{2}\Theta \left(
\vec{\nabla} \phi \right) ^{2}+\left( \vec{\nabla}\Theta \right)
^{2}\right) +{C}_{1}(\phi ),  \label{eq:Func11}
\end{eqnarray}
\begin{equation}
F_{2}=\int {d}V\int\limits_{0}^{1}{d}{\lambda}\enskip \phi
{R}_{2}({\lambda} \phi )=\int {d}V\left( S\sin \Theta
\frac{\partial \Theta }{\partial t}\phi -\frac{JS^{2}}{ 2\hbar
}\sin ^{2}\Theta \left( \vec{\nabla}\phi \right) ^{2}\right)
+{C}_{2}(\Theta ).  \label{eq:Func2}
\end{equation}
\smallskip The ${C}_{1,2}$ are some functions which are determined by the
condition $F_{1}=F_{2}=F(\Theta ,\phi )$. The using of the standard
parametrization
\begin{equation}
\vec{S}({\lambda})=S\vec{N}({\lambda}),\;\vec{N}({\lambda})=\left( \sin
({\lambda} \Theta )\cos (\phi ),\sin ({\lambda} \Theta )\sin (\phi ),\cos
({\lambda} \Theta )\right),  \label{eq:param}
\end{equation}
where $\vec{N}(1)=\vec{N}(t)$ is the physical field and $\vec{N}(0)=(0,0,1)$
is the north pole of the physical sphere yields to the Berry phase term in
(\ref{eq:Func11})
\begin{eqnarray}
S_{{B}} &=&\int {d}V\int\limits_{0}^{1}{d}{\lambda} (-S\sin
({\lambda}\Theta ))\frac{
\partial ({\lambda} \Theta )}{\partial {\lambda}}\frac{\partial \phi }{\partial
t}  \nonumber \\
&=&-S\int
{d}\vec{r}\int\limits_{t_{1}}^{t_{2}}{d}t\int\limits_{0}^{1}{d}{\lambda}
\left( \vec{N}\left[ \frac{\partial \vec{N}}{\partial
{\lambda}}\times \frac{
\partial \vec{N}}{\partial t}\right] \right) .  \label{eq:Berry}
\end{eqnarray}
The comparison of Eqs. (\ref{eq:Func11}, \ref{eq:Func2}) gives the
condition $\phi (t_{1})=\phi (t_{2})$. In contrary to a path
integral formulation the claim $\vec{N}(t_{1})= \vec{N}(t_{2})$
arises as a necessary condition of potentiality of the
differential operator. The following surface integrals must be
zero $\;S_{1}=\oint\limits_{S}{d}\vec{S}\Theta \vec{\nabla}\Theta
$ and $S_{2}=\oint\limits_{S}{d}\vec{S}\phi \sin ^{2}\Theta
\vec{\nabla}\phi$. The unifying of Eqs.(\ref{eq:Func11},
\ref{eq:Func2}) using the divergence theorem  gives the final form
of the functional
\begin{equation}
F=S_{{B}}-\frac{1}{\hbar }\int
{d}\vec{r}\int\limits_{t_{1}}^{t_{2}}{d}t\left \langle \Omega
\left| H\right| \Omega \right\rangle ,  \label{eq:action}
\end{equation}
where the part
\begin{equation}
\left\langle \Omega \left| H\right| \Omega \right\rangle =-JS^{2}\left( n-
\frac{1}{2}\sin ^{2}\Theta \left( \vec{\nabla}\phi \right) ^{2}-\frac{1}{2}
\left( \vec{\nabla}\Theta \right) ^{2}\right)   \label{eq:energy}
\end{equation}
is the density of the energy of the system. The expression (\ref{eq:action})
presents the long-vawe action (Euclidean) if to replace $t\rightarrow {i} \tau
$ ($\tau \in [0,1/T]$) and multiply the functional by $-{i}$.

The same analysis can be made for a generic Heisenberg two-sublattice
antiferromagnet ($J<0$) with  the non-linear differential equations forming a vector of differential operators
\begin{eqnarray}
{R}_{1}&=&-S\sin \Theta _{1}\frac{\partial \phi _{1}}{\partial t}+\frac{%
JS^{2}}{\hbar }\cos \Theta _{1}\sin \Theta _{2}\cos (\phi _{1}-\phi
_{2})\left( n-\left( \vec{\nabla}\Theta _{2}\right) ^{2}-\left( \vec{\nabla}%
\phi _{2}\right) ^{2}\right) \nonumber \\
& &+\frac{JS^{2}}{\hbar }\cos \Theta _{1}\sin \Theta _{2}\sin (\phi
_{1}-\phi _{2})\bigtriangleup \phi _{2}
+\frac{JS^{2}}{\hbar }\cos \Theta _{1}\cos \Theta _{2}\cos (\phi
_{1}-\phi _{2})\bigtriangleup \Theta _{2} \nonumber  \\
& &+\frac{2JS^{2}}{\hbar }\cos \Theta _{1}\cos \Theta _{2}\sin (\phi
_{1}-\phi _{2})\left( \vec{\nabla}\phi _{2}\vec{\nabla}\Theta
_{2}\right)
+\frac{JS^{2}}{\hbar }\sin \Theta _{1}\sin \Theta _{2}\bigtriangleup \Theta
_{2} \nonumber  \\
& &-\frac{JS^{2}}{\hbar }\sin \Theta _{1}\cos \Theta _{2}\left( n-\left( \vec{%
\nabla}\Theta _{2}\right) ^{2}\right)=0, \
\end{eqnarray}
${R}_{2}$ is the analogous equation with substitution of the lower indexes $1 \rightleftharpoons 2$,
\begin{eqnarray}
{R}_{3}&=&S\sin \Theta _{1}\frac{\partial \Theta _{1}}{\partial t}+\frac{JS^{2}%
}{\hbar }\sin \Theta _{1}\sin \Theta _{2}\sin (\phi _{2}-\phi
_{1})\left( n-\left( \vec{\nabla}\Theta _{2}\right) ^{2}-\left( \vec{\nabla}%
\phi _{2}\right) ^{2}\right) \nonumber  \\
& &+\frac{JS^{2}}{\hbar }\sin \Theta _{1}\cos \Theta _{2}\sin (\phi
_{2}-\phi _{1})\bigtriangleup \Theta _{2}
+\frac{JS^{2}}{\hbar }\sin \Theta _{1}\sin \Theta _{2}\cos (\phi
_{1}-\phi _{2})\bigtriangleup \phi _{2} \nonumber  \\
& &+\frac{2JS^{2}}{\hbar }\sin \Theta _{1}\cos \Theta _{2}\cos (\phi
_{1}-\phi _{2})\left( \vec{\nabla}\Theta _{2}\vec{\nabla}\phi
_{2}\right)=0 \
\end{eqnarray}
and the same equation ${R}_{4}$ with $1 \rightleftharpoons 2$.
  Here, one must suppose $z_{1}=\Theta _{1},z_{2}=\Theta
_{2},z_{3}=\phi _{1},z_{4}=\phi _{2}$ where the lower indexes in
$\Theta $ and $\phi $ denote the sublattices. The symmetry test
$\int u_{1}{R}_{,3}^{1}u_{3}{d}V=\int
u_{3}{R}_{,1}^{3}u_{1}{d}V$\smallskip and $\int
u_{2}{R}_{,4}^{2}u_{4}{d}V=\int u_{4}{R}_{,2}^{4}u_{2}{d}V$ gives
the time boundary conditions $\sin \Theta _{1}(t_{1})=\sin \Theta
_{1}(t_{2})$ and $\sin \Theta _{2}(t_{1})=\sin \Theta
_{2}(t_{2}),$ correspondingly. The rest symmetry conditions
determine the surface integrals which must be zero if the
differential operator is potential. The construction of the
variational principle gives
\begin{eqnarray}
F_{1}=\int {d}V \int\limits_{0}^{1} {d}\lambda \Theta _{1}
{R}_{1}(\lambda \Theta _{1}) =
-S\int {d}V\int\limits_{0}^{1}{d}\lambda \left( \vec{N}_{1}\left[ \frac{%
\partial \vec{N}_{1}}{\partial \lambda }\times \frac{\partial \vec{N}_{1}}{%
\partial t}\right] \right) \nonumber  \\
+\frac{JS^{2}}{\hbar} \int {d}V  \sin \Theta _{1}\left( \sin
\Theta _{2}\cos (\phi _{1}-\phi
_{2})\left( n-\left( \vec{\nabla}\phi _{2}\right) ^{2}-\left( \vec{\nabla}%
\Theta _{2}\right) ^{2}\right) \right. \nonumber  \\
+\cos {\Theta}_{2} \cos ({\phi}_{1}-{\phi}_{2}) \Delta \Theta
_{2}+\sin {\Theta}_{2} \sin ({\phi}_{1}-{\phi}_{2})\Delta {\phi}_{2} \nonumber  \\
\left. +2\cos \Theta _{2}\sin (\phi _{1}-\phi _{2})\left( \vec{\nabla}%
\Theta _{2}\cdot \vec{\nabla}\phi _{2}\right) \right) \nonumber  \\
+\frac{JS^{2}}{\hbar} \int {d}V\cos \Theta _{1}\left( \cos \Theta _{2}\left( n-\left( \vec{\nabla}%
\Theta _{2}\right) ^{2}\right) -\sin \Theta _{2}\Delta \Theta _{2}\right)+
{C}_{1}(\phi_{1},\Theta _{2},\phi _{2}) \nonumber  \\
\
\end{eqnarray}
and
\begin{eqnarray}
F _{2}=\int {d}V \int\limits_{0}^{1} {d}\lambda \phi _{1}
{R}_{3}(\lambda \phi _{1})=
-S\int {d}V\int\limits_{0}^{1}{d}\lambda \left( \vec{N}_{1}\left[ \frac{%
\partial \vec{N}_{1}}{\partial \lambda }\times \frac{\partial \vec{N}_{1}}{%
\partial t}\right] \right) \nonumber  \\
+\frac{JS^{2}}{\hbar} \int {d}V \sin \Theta _{1}\left( \sin \Theta
_{2}\cos (\phi _{1}-\phi
_{2})\left( n-\left( \vec{\nabla}\phi _{2}\right) ^{2}-\left( \vec{\nabla}%
\Theta _{2}\right) ^{2}\right) \right. \nonumber  \\
+\cos {\Theta}_{2} \cos ({\phi}_{1}-{\phi}_{2}) \Delta \Theta
_{2}+\sin {\Theta}_{2} \sin ({\phi}_{1}-{\phi}_{2})\Delta {\phi}_{2} \nonumber  \\
\left. +2\cos \Theta _{2}\sin (\phi _{1}-\phi _{2})\left( \vec{\nabla}%
\Theta _{2}\cdot \vec{\nabla}\phi _{2}\right) \right) +{C}_{2}(\Theta
_{1},\Theta _{2},\phi _{2}),  \
\end{eqnarray}
where ${C}_{1,2}$ are some functions. The comparison of both forms yields to $F(\Theta _{1},\phi _{1}) + {C}_{3}(\Theta _{2},\phi _{2})=F_{1}=F_{2}$  if the following conditions are fulfiled
\begin{equation}
{C}_{1}(\phi_{1},\Theta _{2},\phi _{2})=
{C}_{3}(\Theta _{2},\phi _{2}), \label{eq:C3}
\end{equation}
\begin{eqnarray}
{C}_{2}(\Theta_{1},\Theta _{2},\phi _{2})={C}_{3}(\Theta _{2},\phi _{2}) \nonumber  \\
+\frac{JS^{2}}{\hbar} \int {d}V\cos \Theta _{1}\left( \cos \Theta
_{2}\left( n-
\left( \vec{\nabla}%
\Theta _{2}\right) ^{2}\right) -\sin \Theta _{2}\Delta \Theta _{2}\right)
. \  \label{eq:C1}
\end{eqnarray}
The similar expression may be obtained for $F(\Theta _{2},\phi _{2})$. It equals to
$F(\Theta _{1},\phi _{1})$ if to replace $1 \rightleftharpoons 2$. The final form for the variational principle may be obtained via an integration by parts using the divergence theorem and the boundary conditions found at the symmetry analysis. The result is
\begin{equation}
F=-S\sum\limits_{i=1}^{2}\int
{d}\vec{r}\int\limits_{t_{1}}^{t_{2}}{d}t\int
\limits_{0}^{1}{d}{\lambda} \left( \vec{N}_{i}\left[
\frac{\partial \vec{N}_{i}}{
\partial {\lambda} }\times \frac{\partial \vec{N}_{i}}{\partial t}\right]
\right) -\frac{1}{\hbar }\int
{d}\vec{r}\int\limits_{t_{1}}^{t_{2}} {d}t\left \langle \Omega
\left| H\right| \Omega \right\rangle .  \label{eq:AFM}
\end{equation}

As in the previous case the expression coincides with the action and
includes the Berry phase terms from each sublattices. The density of the
energy is given as following
\begin{eqnarray}
\left\langle \Omega \left| H\right| \Omega \right\rangle & = & JS^{2}(-n\sin
\Theta _{1}\sin \Theta _{2}\cos (\phi _{1}-\phi _{2})-n\cos \Theta
_{1}\cos \Theta _{2}  \nonumber \\
& &+\cos \Theta _{1}\cos \Theta _{2}\cos (\phi _{1}-\phi _{2})\left(
\vec{\nabla}\Theta _{1}\vec{\nabla}\Theta _{2}\right) \nonumber \\
& &-\sin \Theta _{1}\cos \Theta _{2}\sin (\phi _{1}-\phi _{2})\left(
\vec{\nabla}\phi _{1}\vec{\nabla}\Theta _{2}\right)  \nonumber \\
& &+\cos \Theta _{1}\sin \Theta _{2}\sin (\phi _{1}-\phi _{2})\left(
\vec{\nabla}\Theta _{1}\vec{\nabla}\phi _{2}\right)  \nonumber \\
& &+\sin \Theta _{1}\sin \Theta _{2}\cos (\phi _{1}-\phi _{2})\left(
\vec{\nabla}\phi _{1}\vec{\nabla}\phi _{2}\right) \nonumber \\
& &+\sin \Theta _{1}\sin \Theta _{2}\left( \vec{\nabla}\Theta _{1}\vec{\nabla}
\Theta _{2}\right)).  \label{eq:AFMen}
\end{eqnarray}
In conclusion, the method of derivation of the long-wave action for spin
Hamiltonians is suggested via the solving of the inverse problem of calculus
variations. The case of a Heisenberg (anti)ferromagnet is considered.

\end{document}